\documentclass[12pt]{cernart}
\usepackage[latin1]{inputenc}
\usepackage[dvips]{graphicx,epsfig,color}
\usepackage{wrapfig,rotating}
\usepackage{amssymb}
\usepackage{cite}

\def\lapproxeq{\lower .7ex\hbox{$\;\stackrel{\textstyle
<}{\sim}\;$}}
\def\gapproxeq{\lower .7ex\hbox{$\;\stackrel{\textstyle
>}{\sim}\;$}}

\newcommand{\Deqt }{$\Delta_T q$}
\newcommand{\Deqx }{$\Delta q(x)$}
\newcommand{\Deqtx }{$\Delta_T q(x)$}
\newcommand{\qx }{$q(x)$}

\newcommand{\AmS}{{\protect\the\textfont2
  A\kern-.1667em\lower.5ex\hbox{M}\kern-.125emS}}

\begin {document}
\dimen\footins=\textheight

\begin{titlepage}
\docnum{CERN--PH--EP/2008--002}
\hbox to \hsize{\hskip123mm\hbox{5 February 2008}\hss}
\hbox to \hsize{\hskip123mm\hbox{revised 16 December 2008}\hss}
\vspace{1cm}

\begin{center}
{\LARGE {\bf Collins and Sivers asymmetries for pions and kaons in muon--deuteron DIS}}
\vspace*{0.5cm}
\end{center}

\author{\large The COMPASS Collaboration}

\vspace{2cm}
\begin{abstract}
The measurements of the Collins and Sivers asymmetries 
of identified hadrons produced 
in deep-inelastic scattering of 160~GeV/$c$ muons on 
a transversely polarised $^6$LiD 
target at COMPASS are presented. 
The results for charged pions and charged and neutral kaons 
correspond to all data available, which were collected
from 2002 to 2004.
For all final state particles both, the Collins and Sivers asymmetries  
turn out to be small, compatible with zero within the 
 statistical errors, in line with the previously published 
results for not identified charged hadrons, and with the expected
cancellation between the u- and d-quark contributions.
\end{abstract}

\vspace*{60pt}
\noindent
Keywords: 
transversity, deuteron, transverse single-spin asymmetry, 
identified hadrons, Collins asymmetry, Sivers asymmetry, COMPASS.\\

\noindent
PACS 13.60.-r, 13.88.+e, 14.20.Dh, 14.65.-q
\vfill
\submitted{Physics Letters B}

%
%
\begin{Authlist}
{\large  The COMPASS Collaboration}\\[\baselineskip]
{\small
%
%
M.~Alekseev\Iref{turin_p},
V.Yu.~Alexakhin\Iref{dubna},
Yu.~Alexandrov\Iref{moscowlpi},
G.D.~Alexeev\Iref{dubna},
A.~Amoroso\Iref{turin_u},
A.~Arbuzov\Iref{dubna},
B.~Bade\l ek\Iref{warsaw},
F.~Balestra\Iref{turin_u},
J.~Ball\Iref{saclay},
J.~Barth\Iref{bonnpi}
% ...
}
\end{Authlist}

%
%
\Instfoot{bielefeld}{Universit\"at Bielefeld, Fakult\"at f\"ur Physik, 33501 Bielefeld, Germany\Aref{e}}
\Instfoot{bochum}{Universit\"at Bochum, Institut f\"ur Experimentalphysik, 44780 Bochum, Germany\Aref{e}}
\Instfoot{bonniskp}{Universit\"at Bonn, Helmholtz-Institut f\"ur  Strahlen- und Kernphysik, 53115 Bonn, Germany\Aref{e}}
\Instfoot{bonnpi}{Universit\"at Bonn, Physikalisches Institut, 53115 Bonn, Germany\Aref{e}}
\Instfoot{brno}{Institute of Scientific Instruments, AS CR, 61264 Brno, Czech Republic\Aref{f}}
\Instfoot{burdwan}{Burdwan University, Burdwan 713104, India\Aref{g}}
\Instfoot{calcutta}{Matrivani Institute of Experimental Research \& Education, Calcutta-700 030, India\Aref{h}}
\Instfoot{dubna}{Joint Institute for Nuclear Research, 141980 Dubna, Moscow region, Russia}
\Instfoot{erlangen}{Universit\"at Erlangen--N\"urnberg, Physikalisches Institut, 91054 Erlangen, Germany\Aref{e}}
\Instfoot{freiburg}{Universit\"at Freiburg, Physikalisches Institut, 79104 Freiburg, Germany\Aref{e}}
\Instfoot{cern}{CERN, 1211 Geneva 23, Switzerland}
\Instfoot{heidelberg}{Universit\"at Heidelberg, Physikalisches Institut,  69120 Heidelberg, Germany\Aref{e}}
\Instfoot{helsinki}{Helsinki University of Technology, Low Temperature Laboratory, 02015 HUT, Finland  and University of Helsinki, Helsinki Institute of  Physics, 00014 Helsinki, Finland}
\Instfoot{liberec}{Technical University in Liberec, 46117 Liberec, Czech Republic\Aref{f}}
\Instfoot{lisbon}{LIP, 1000-149 Lisbon, Portugal\Aref{i}}
\Instfoot{mainz}{Universit\"at Mainz, Institut f\"ur Kernphysik, 55099 Mainz, Germany\Aref{e}}
\Instfoot{miyazaki}{University of Miyazaki, Miyazaki 889-2192, Japan\Aref{j}}
\Instfoot{moscowlpi}{Lebedev Physical Institute, 119991 Moscow, Russia}
\Instfoot{munichlmu}{Ludwig-Maximilians-Universit\"at M\"unchen, Department f\"ur Physik, 80799 Munich, Germany\AAref{e}{k}}
\Instfoot{munichtu}{Technische Universit\"at M\"unchen, Physik Department, 85748 Garching, Germany\AAref{e}{k}}
\Instfoot{nagoya}{Nagoya University, 464 Nagoya, Japan\Aref{j}}
\Instfoot{praguecu}{Charles University, Faculty of Mathematics and Physics, 18000 Prague, Czech Republic\Aref{f}}
\Instfoot{praguectu}{Czech Technical University in Prague, 16636 Prague, Czech Republic\Aref{f}}
\Instfoot{protvino}{State Research Center of the Russian Federation, Institute for High Energy Physics, 142281 Protvino, Russia}
\Instfoot{saclay}{CEA DAPNIA/SPhN Saclay, 91191 Gif-sur-Yvette, France}
\Instfoot{telaviv}{Tel Aviv University, School of Physics and Astronomy, 69978 Tel Aviv, Israel\Aref{l}}
\Instfoot{triest_i}{Trieste Section of INFN, 34127 Trieste, Italy}
\Instfoot{triest}{University of Trieste, Department of Physics and Trieste Section of INFN, 34127 Trieste, Italy}
\Instfoot{triestictp}{Abdus Salam ICTP and Trieste Section of INFN, 34127 Trieste, Italy}
\Instfoot{turin_u}{University of Turin, Department of Physics and Torino Section of INFN, 10125 Turin, Italy}
\Instfoot{turin_i}{Torino Section of INFN, 10125 Turin, Italy}
\Instfoot{turin_p}{University of Eastern Piedmont, 1500 Alessandria,  and Torino Section of INFN, 10125 Turin, Italy}
\Instfoot{warsaw}{So{\l}tan Institute for Nuclear Studies and Warsaw University, 00-681 Warsaw, Poland\Aref{m} }
\Instfoot{warsawtu}{Warsaw University of Technology, Institute of Radioelectronics, 00-665 Warsaw, Poland\Aref{n} }
\Instfoot{yamagata}{Yamagata University, Yamagata, 992-8510 Japan\Aref{j} }
%
%
\Anotfoot{+}{Deceased}
\Anotfoot{a}{Also at IST, Universidade T\'ecnica de Lisboa, Lisbon, Portugal}
\Anotfoot{b}{On leave of absence from JINR Dubna}
\Anotfoot{c}{Also at Chubu University, Kasugai, Aichi, 487-8501 Japan}
\Anotfoot{d}{Also at GSI mbH, Planckstr.\ 1, D-64291 Darmstadt, Germany}
\Anotfoot{e}{Supported by the German Bundesministerium f\"ur Bildung und Forschung}
\Anotfoot{f}{Suppported by Czech Republic MEYS grants ME492 and LA242}
\Anotfoot{g}{Supported by DST-FIST II grants, Govt. of India}
\Anotfoot{h}{Supported by  the Shailabala Biswas Education Trust}
\Anotfoot{i}{Supported by the Portuguese FCT - Funda\c{c}\~ao para a Ci\^encia e Tecnologia grants POCTI/FNU/49501/2002 and POCTI/FNU/50192/2003}
\Anotfoot{j}{Supported by the Ministry of Education, Culture, Sports, Science and Technology, Japan, Grant-in-Aid for Specially Promoted Research No.\ 18002006; Daikou Foundation and Yamada Foundation}
\Anotfoot{k}{Supported by the DFG cluster of excellence `Origin and Structure of the Universe' (www.universe-cluster.de)}
\Anotfoot{l}{Supported by the Israel Science Foundation, founded by the Israel Academy of Sciences and Humanities}
\Anotfoot{m}{Supported by Ministry of Science and Higher Education grant 41/N-CERN/2007/0 and the MNII research funds for 2005--2007}
\Anotfoot{n}{Supported by KBN grant nr 134/E-365/SPUB-M/CERN/P-03/DZ299/2000}
%

%
%
\end{titlepage}


%
\section{Introduction}
To fully specify the quark structure of the nucleon at the twist-two
level, the
transverse spin distributions \Deqtx\ must be added to the better known spin-average
distributions \qx\ and to the helicity distributions \Deqx ~\cite{Jaffe:1991kp}.
Here $x$ is the Bjorken variable,
which represents the momentum fraction of the quarks inside the nucleon.
The interpretation of the transversity distribution is similar to that of
the helicity distribution, i.e.\ in a transversely polarised nucleon
\Deqt\ is the difference of the number density of quarks with momentum
fraction $x$ and spin parallel or antiparallel to the nucleon spin.
For a discussion on the notation see Refs.~\cite{Barone:2001sp}
and~\cite{Bacchetta:2004jz}.

\begin{figure}[tb!]
\begin{center}
\includegraphics[width=.70\textwidth]{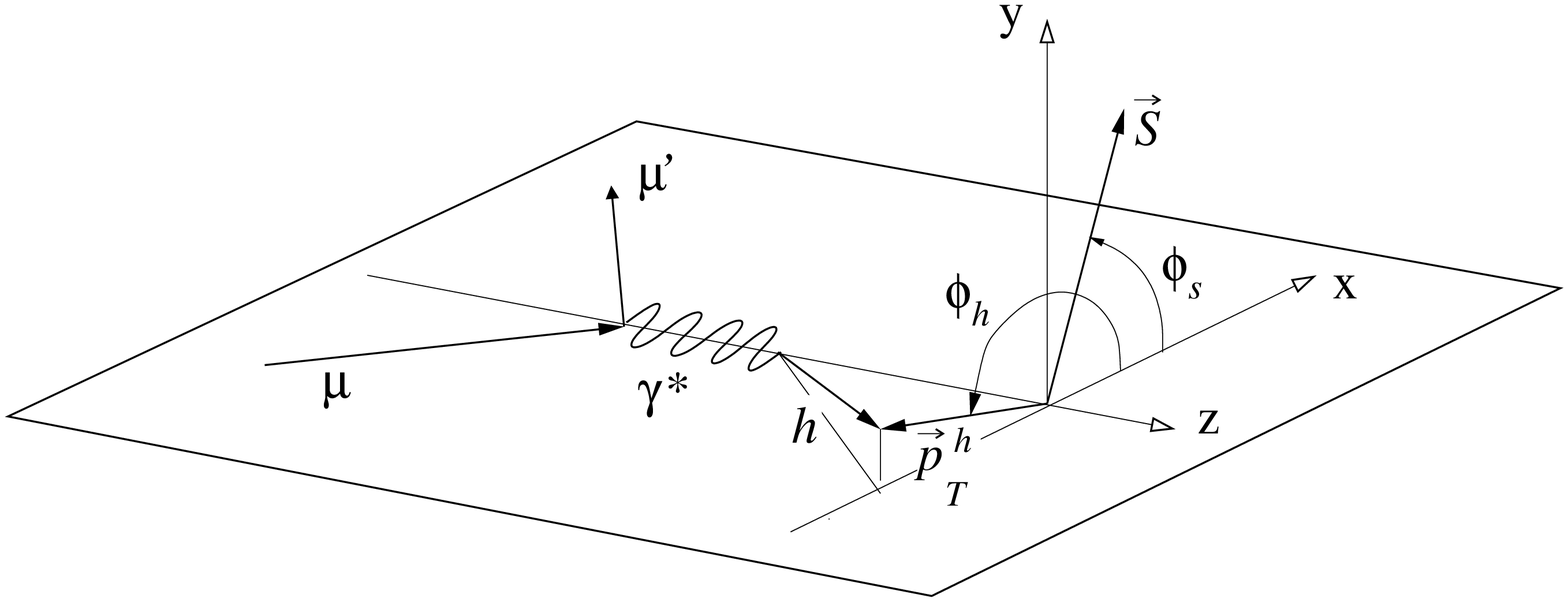} 
\end{center}
\caption{Definition of the azimuthal angle $\phi_h $ of the transverse 
momentum $\vec{p}_T^{\, h}$ of the outgoing hadron  and of 
the azimuthal angle  $\phi_{S}$  of
the transverse spin vector $\vec{S}$ of the target nucleon.
}
\label{fig:gns}
\end{figure}

The distributions \Deqt\ are difficult to measure, since they are chirally odd
and therefore absent in inclusive deep inelastic scattering (DIS). 
They may instead be extracted from
measurements of the single-spin azimuthal asymmetries in cross-sections for
semi-inclusive DIS (SIDIS) of leptons off transversely polarised nucleons,
in which a hadron is also detected in the final state. 
In these processes
the measurable asymmetry is due to the combined effect of \Deqt\ and
a chirally-odd fragmentation function (FF) which describes the spin-dependent part of
the hadronization of a transversely polarised quark. 
At leading twist, the existence of 
such a naively
$T$-odd FF arising from final state interaction
effects, was predicted by Collins~\cite{Collins:1992kk} and is now generally known
as the Collins effect. 
In the
fragmentation of transversely polarised quarks
it is responsible for a left-right asymmetry which is
due to a correlation between the spin of the fragmenting quark 
and the transverse momentum $\vec{p}_{\perp}$ of the produced hadron with
respect to the quark direction.
The $\vec{p}_{\perp}$-dependent fragmentation function 
of a transversely polarised quark $q$ into a
spinless hadron $h$ is thus expected to be of the form
\begin{equation}
D_{T\,q}^{\; \; \; h}(z, \vec{p}_{\perp}) = D_q^h(z, {p}_{\perp}) + 
    \Delta_T^0 D_q^h(z, {p}_{\perp}) \cdot \sin\varphi \, ,
\label{eq:collfun}
\end{equation}
where $D_q^h$ is the unpolarised FF and the ``Collins function''
$\Delta_T^0 D_q^h$ is the 
$T$-odd part of the FF,
responsible for the left-right asymmetry,
and $z$ is the fraction of available energy carried by the hadron.
Here $\varphi$ is the difference of the azimuthal angles of the hadron transverse 
momentum  and the
quark spin,  relative to the quark direction.
As a result, in SIDIS off transversely polarised nucleons the Collins
mechanism is responsible for a modulation in the azimuthal distribution
of the  produced  hadrons given by
\begin{equation}
N(\Phi_C)= N_0 (1+\epsilon_{C} \cdot \sin\Phi_C) \, .
\label{eq:colldis}
\end{equation}
In a gamma--nucleon reference system (GNS), in which 
the $z$-axis coincides with the virtual photon direction
and the $x$-$z$ plane is the lepton scattering plane,
 the ``Collins angle'' $\Phi_C$ is $\Phi_C=\phi_h + \phi_{S} - \pi$.
Here $\phi_h $ is
the azimuthal angle of the transverse momentum $\vec{p}_T^{\, h}$ of the outgoing hadron 
and $\phi_{S}$  is
the azimuthal angle  of
the transverse spin vector $\vec{S}$ of the target nucleon, as
shown in Fig.~\ref{fig:gns}.
The measurable asymmetry $\epsilon_{C}$ is related to the convolution of the
transversity parton distribution function (PDF)  \Deqt\ and the 
Collins function.

An entirely different mechanism was suggested by Sivers~\cite{Sivers:1989cc} 
as a possible cause of the transverse spin effects observed in $pp$ scattering.
This mechanism could also be responsible for a spin asymmetry in the 
cross-section of SIDIS of leptons off
transversely polarised nucleons. 
Sivers' conjecture was based on a possible
existence of a correlation between the intrinsic transverse momentum
$\vec{k}_{\perp}$ of a quark and the
transverse polarisation vector of the nucleon $\vec{S}$, 
i.e.\ that the quark distribution $q(x)$ could
be written as
\begin{equation}
q_{T}(x,\vec{k}_{\perp})= q(x,{k}_{\perp}) + | \vec{S} | \cdot 
            \Delta_0^T q(x,{k}_{\perp})\cdot \sin \varphi ',
  \label{eq:fsiv}
\end{equation}
where $\varphi '$ is the difference of the azimuthal angles 
of the transverse spin of the nucleon and of the quark  transverse momentum,
 relative to the nucleon direction.
In SIDIS off transversely polarised nucleons the Sivers
mechanism results in a modulation in the azimuthal distribution
of the  produced hadrons
\begin{equation}
N(\Phi_S)= N_0 (1+\epsilon_{S} \cdot \sin\Phi_S) \, ,
\label{eq:sivdis}
\end{equation}
where the ``Sivers angle" $\Phi_{ S}= \phi_h -\phi_S$
is the relative azimuthal angle between the transverse momentum of the hadron $p_T^h$
and the nucleon target spin in the GNS.
In this case, the measurable asymmetry $\epsilon_{S}$ is related to the convolution of the
Sivers PDF $\Delta_0^Tq$ and the unpolarised FF $ D_q^h$.

Since the Collins and Sivers terms in the transverse spin asymmetry depend on
the two independent angles $\Phi_C$ and $\Phi_S$, measuring SIDIS on a transversely
polarised target allows the Collins and the Sivers effects to be
disentangled, and the two asymmetries can separately be extracted from the
data.
The same is true for the other azimuthal asymmetries which appear
in the general expression of the SIDIS cross-section in the one photon
exchange approximation~\cite{Bacchetta:2006tn}. 
Correlations between the different terms 
can  be introduced by a
non-constant acceptance of the apparatus.

Collins and Sivers  modulations 
have been shown experimentally to be non
zero by the HERMES measurements of pion asymmetries in SIDIS
on a proton target~\cite{Airapetian:2004tw,Diefenthaler:2007rj}.
Independent information on the Collins function has been provided by the
azimuthal correlations in $e^+e^-\rightarrow$ hadrons measured by the BELLE
Collaboration~\cite{Abe:2005zx}.
COMPASS has already published results for the Collins and Sivers asymmetries 
with a deuteron target for non-identified hadrons~\cite{Alexakhin:2005iw,Ageev:2006da},
and the details of the experimental
technique can be found there.
Here we present the Collins and Sivers asymmetries on a transversely
polarised deuteron target for identified hadrons, i.e.\ charged pions,
and charged and neutral kaons, which  put more stringent constraints on a flavour separated
analysis of these new transverse spin effects.

\section{The COMPASS experiment and the SIDIS event selection}
\label{expSect}
The COMPASS experiment is set up at the M2 beam line at 
CERN, using both muon and hadron beams.  
For this measurement a longitudinally polarised $\mu^+$ beam of 160~GeV/$c$ momentum was
 scattered off a solid $^6$LiD polarised target consisting of two cylindrical cells 
along the beam direction.  
The two cells were polarised in opposite directions, so that 
data were taken simultaneously on oppositely polarised targets 
to reduce the systematic errors.
The target polarisation direction could be set either parallel or orthogonal to 
the beam direction.

Particle tracking and identification are performed in a two-stage 
spectrometer, covering a wide kinematical range.  
A Ring Imaging Cherenkov detector (RICH-1)
and two hadron calorimeters provide particle identification.  
The RICH-1~\cite{rich1} detector is a gas RICH with a 3~m long C$_4$F$_{10}$
radiator 
covering the whole spectrometer acceptance. 
Two spherical mirror surfaces 
reflect and focus the Cherenkov photons on two sets of detectors
 outside the acceptance region. The photon detection utilises 
multiwire proportional chambers
(MWPC) with segmented CsI photocathodes which detect photons 
in the UV region.
The trigger system comprises hodoscope counters and hadron calorimeters.  
Veto counters installed in front of the target are used to reject the beam halo.  
A detailed description of the spectrometer can be found in Ref.~\cite{Abbon:2007pq}.

The COMPASS Collaboration has taken data with the $^6$LiD target 
polarised transversely with respect to the incoming beam direction 
in 2002, 2003, and 2004. 
The transversity data were taken
for about 20\% of the available beam time.  
During data taking, particular care was taken to ensure the stability of 
the apparatus.  
The target polarisation was typically reversed every 5 days 
to reduce systematic effects due to the different acceptances of the two cells.  
To guarantee that the acceptance of the detector fulfills the stability 
requirements the distributions of several physical quantities 
($x$, $\vec{p}_T^{\, h}$, \ldots)  
have been monitored.

In the data analysis, the selection of events required
a ``primary vertex'', defined by the incoming and the scattered
muon tracks, and at least one hadron outgoing from the 
primary vertex.
In addition  the projected beam track was required to cross both target cells. 
Clean hadron and muon selection was achieved using the hadron calorimeters 
and considering the amount of traversed material.
To select DIS events, cuts on the photon virtuality $Q^2>1~($GeV/$c)^2$ and  the mass 
of the hadronic final  state $W>5$~GeV/$c^2$ were applied. 
The requirement of $0.1 < y < 0.9 $,
where $y$ is the fractional energy of the virtual photon, 
limits the error due to radiative corrections 
and avoids contamination from $\pi$ decay (upper cut), and warrants 
a good determination of $y$ (lower cut).
To safely reconstruct the hadron azimuthal angle $\phi_h$ a 
minimum transverse momentum $p_T^h$ 
 of 0.1~GeV/$c$ is required.
Furthermore, to select the current fragmentation region, a 
lower limit for   the relative 
energy of the hadron $z$ is 
required.
In the following, all hadrons with $z>0.2$ define the ``all hadron'' sample.
The analysis is performed also for ``leading 
hadrons'' only, for which we expect an enhancement of the
physics signal albeit lower statistics.
The ``leading hadron'' is defined as the most energetic hadron 
coming from the primary vertex
and having $z>0.25$.
Detailed information on the data analysis can be found in Ref.~\cite{Ageev:2006da}.
Additional cuts specific to the analysis of the identified charged and neutral hadrons
 are described in the next sections.

\section{Charged particle identification}
\label{sectionWithRich}

Charged hadrons are identified as $\pi$ and $K$ using the 
 RICH-1 detector.
 Several variables have been monitored to ensure the stability in 
time of the  RICH-1 response.
The monitored variables are the hit multiplicities of the MWPCs,
 the mass hypothesis likelihood (described later in this section), and
 the number of hadrons identified as $\pi$ and $K$ normalised to the 
number of reconstructed tracks.
Blocks of data are rejected if any of 
the variables deviates from the mean values by more than 
three    standard deviations.

The identification procedure relies on  a likelihood function 
constructed with the photons detected in RICH-1
and associated to the charged particle trajectory. 
 The likelihood function uses the photons of the signal and
the theoretical expectation from 
the Frank and Tamm equation, taking into account possible signal losses due to dead 
zones in the detector.
  The description of the background photons, 
coming from other particles in the event and from the beam halo,
 is taken from the photon detectors occupancy in the  data.

 Since the number of Cherenkov photons  depends on the velocity 
of the particle, 
for a given momentum the signal yield is different for different mass hypotheses. 
Thus for each track the likelihood is computed for different mass and the 
background hypotheses. 
The particle identification is made by choosing the mass hypothesis corresponding 
to the highest value of the likelihood.
To assure a clear distinction from the background, a cut on the ratio 
of the highest  likelihood 
to the background likelihood is made.
To insure a good separation with respect to  different masses, 
a cut on the ratio of  the  likelihood 
to the second highest likelihood is also made.
The cuts on these variables have been tuned on  subsamples of the  data
 considering  events in which at least two oppositely charged hadrons 
from the primary vertex have been found.
 The cuts for $K$ identification have been tuned on the $\phi$ meson peak in 
the invariant mass 
distribution of the two charged hadrons, 
maximising the product of the $\phi$ signal and the signal-to-background 
ratio  values in the peak.
To tune the cuts for pion identification, we used the same approach,
but using the $\rho$ peak  in the invariant mass distribution.
In addition, all these cuts have been verified on Monte Carlo data.

\begin{figure}[tb]
\begin{center}
\includegraphics[width=0.48\textwidth]{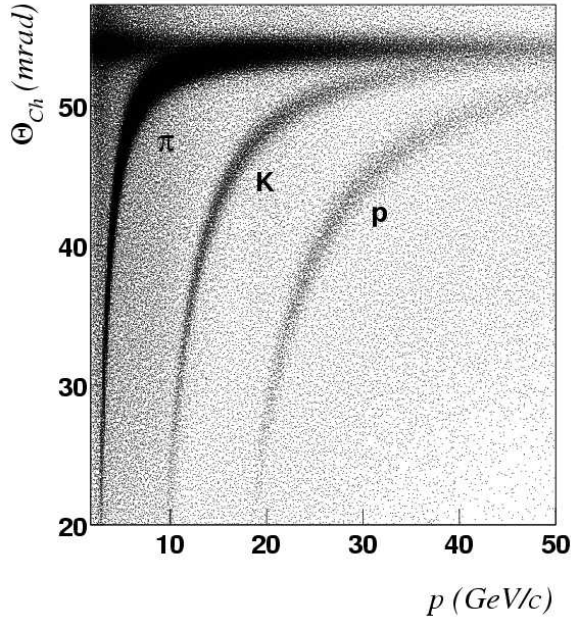}
\end{center}
\caption{The Cherenkov angle
as measured with RICH-1 versus the particle momentum as measured by
the  spectrometer.
}
\label{fig:richmom}
\end{figure}

In Fig.~\ref{fig:richmom} the dependence of 
the Cherenkov angle measured by the RICH-1 against the particle momentum 
clearly indicates the different regions in which hadrons can be identified.  
In particular, the Cherenkov thresholds 
are visible: they are about 2.6~GeV/$c$ for $\pi$, 9~GeV/$c$ for $K$, and 
17~GeV/$c$ for protons.  
To assure a minimum number of detected photons the hadron momenta 
have to be 0.5~GeV/$c$ above threshold for pions 
and 1~GeV/$c$ above threshold for kaons.
The values chosen 
correspond to 4 emitted photons.  
The upper limit for the momentum has been set to 50~GeV/$c$ for both $\pi$ and $K$, 
corresponding 
to 1.5 $\sigma$ mass separation between these two mass hypotheses.

\begin{figure}[tb]
\begin{center}
\includegraphics[width=0.6\textwidth]{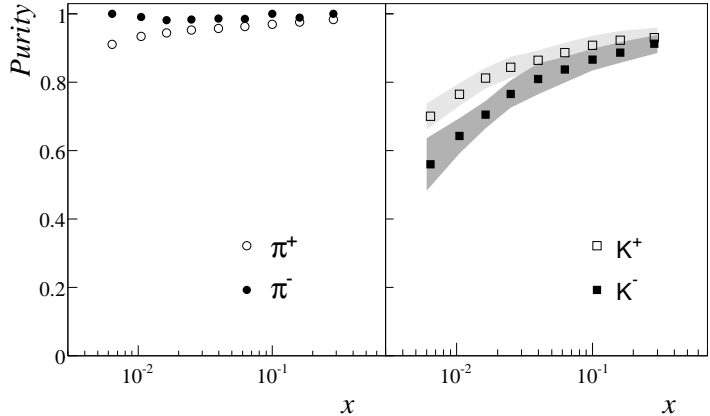}
\end{center}
\caption{Purities of the pion (left) and kaon (right) samples as a function of $x$. 
The open (closed) points are for positive (negative) pions, while open (closed)
squares are for positive (negative) kaons. The shaded area on the right picture
gives the evaluated systematic uncertainty for the kaons, while the same is
negligible for pions.}
\label{fig:purity}
\end{figure}

The purity of the identified hadrons, defined as the fraction of pions
inside the identified pion sample, and the fraction of kaons inside the
identified kaon sample, has been evaluated from the data. The results for this
analysis are shown in Fig.~\ref{fig:purity} for positive and negative pions (left)
and kaons (right) as a  function of $x$.
While the purity for pions is very high and almost independent from $x$, for
kaons the purity depends on the momentum and the polar angle of the hadron, and as a
consequence it increases with $x$.
This trend is less pronounced as a function of $z$ and $p_T$ given the smaller
correlation with the polar angle.  On average, purities are higher than 95\% for
pions, around 70\% for negative kaons and around 80\% for positive kaons.

The final statistics for the different data taking periods are summarised in 
Table~\ref{tab:stat}. The use of the RICH-1 information was not possible for the
2002 data, since the functionality of the detector during the
transversity run in that year was not satisfactory.
\begin{table*}[htb]
\caption{Final statistics for the ``all hadron'' and the ``leading hadrons'' samples.}
\label{tab:stat}
\begin{center}
\vspace*{.5cm}
\begin{tabular}{ l l c c c c c c c c c c }
\hline
  &  & \multicolumn{5}{c}{``all hadrons'' ($\cdot 10^{-6}$)} & \multicolumn{5}{c}{``leading hadrons'' ($\cdot 10^{-6}$)} \\
\hline
Year  & Period & $\pi^+$ &  $\pi^-$ &  $K^+$ &  $K^-$ &  $K_S^0$  & $\pi^+$ &  $\pi^-$ &  $K^+$ &  $K^-$ &  $K_S^0$ \\
\hline
\hline
2002 & 1 & - & - & - & - & 0.021  & - & - & - & - & 0.014\\
2002 & 2 & - & - & - & - & 0.014  & - & - & - & - & 0.009\\
\hline
2003 &   & 1.71 & 1.49 & 0.31 & 0.20 & 0.077  & 1.10 & 0.93 & 0.24 & 0.14 & 0.052 \\
\hline
2004 & 1 & 1.54 & 1.33 & 0.27 & 0.18 & 0.063  & 0.98 & 0.82 & 0.21 & 0.13 & 0.043\\
2004 & 2 & 2.03 & 1.76 & 0.36 & 0.24 & 0.083  & 1.30 & 1.09 & 0.27 & 0.17 & 0.056\\
\hline
\hline
 \multicolumn{2}{l}{Total}  & 5.28 & 4.58 & 0.94 & 0.62 & 0.258 & 3.38 & 2.84 & 0.72  & 0.44 & 0.175 \\
\hline
\end{tabular}\\[2pt]
\end{center}
\vspace*{.8cm}
\end{table*}

\section{$K^0_S$ identification}
The reconstruction of the $K^0_S$ relies on 
the detection of the two decay pions.  
The large acceptance of the COMPASS spectrometer provides a good efficiency for 
the detection of the pion pair.
In this part of the analysis the RICH-1 detector is not 
used to identify the pions.

\begin{figure}[tb!]
\begin{center}
\includegraphics[width=.70\textwidth]{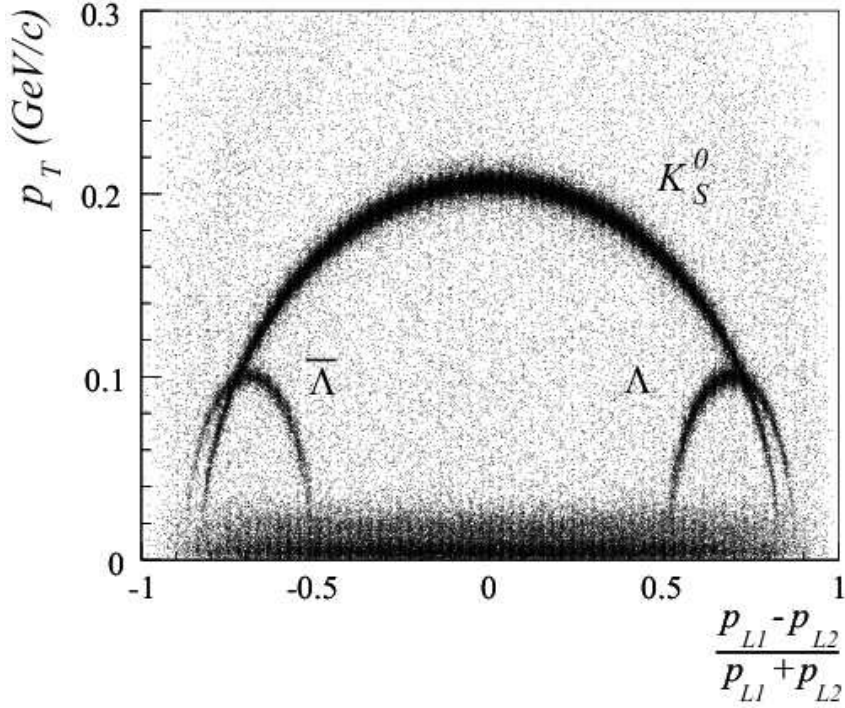} 
\end{center}
\caption{  Armenteros plot of the hadron 
pair. The $K^0_S$ band  can be clearly seen as well as the 
$\Lambda$ and $\bar{\Lambda}$ bands.
}
\label{fig:armenteros}
\end{figure}

\begin{figure}[tb]
\begin{center}
\includegraphics[width=.55\textwidth]{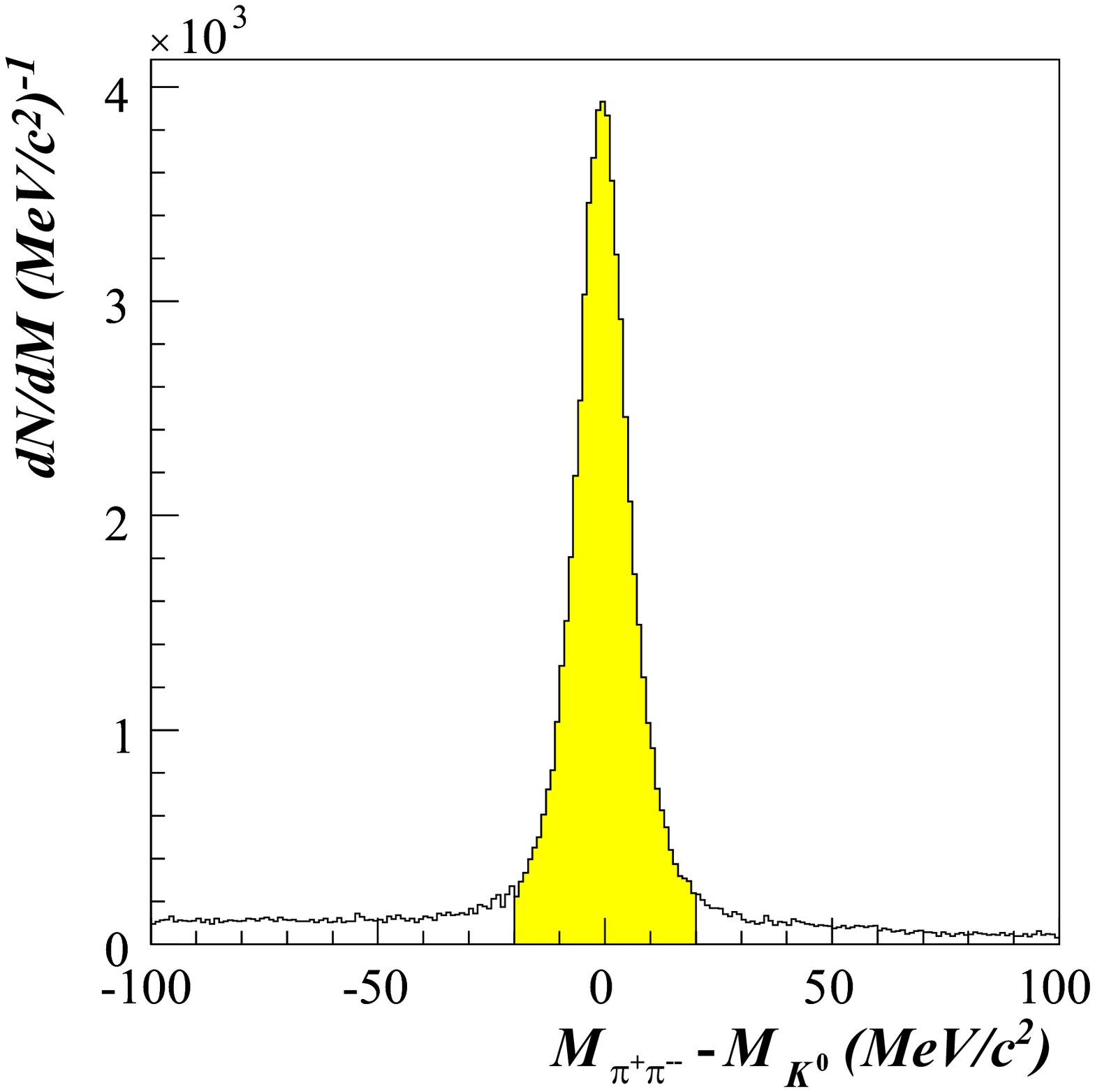} 
\end{center}
\caption{Difference of the invariant mass of the hadron pair 
after cuts to the $K^0_S$ mass. The shaded region shows the
accepted  $K^0_S$.
}
\label{fig:m_inv}
\end{figure}

The signature of $K^0_S$ events is a $V_0$ vertex, 
i.e.\ a vertex with no incoming but two outgoing 
charged particles, where the two detected particles have opposite charge.  
The sum of the two outgoing particle momenta must point to the primary 
vertex and the invariant mass of the two particle system,
assuming pion mass for each of them, must agree with 
the $K^0_S$ mass.
In order to identify $K^0_S$ from the primary vertex, 
all $V_0$ vertices downstream of the primary vertex 
have been considered.  
Moreover, the outgoing
tracks are not allowed to be additionally associated with any primary vertex 
and have to satisfy the criteria of section~2.
To test the association of the secondary vertex to the primary vertex, 
the angle between the reconstructed momentum of the hadron 
pair and the vector connecting the primary and the secondary vertex is 
calculated.  
A maximum angle of 10~mrad is accepted.  
To ensure a proper distinction of the primary and secondary vertices 
a cut on the distance of the primary and the secondary vertex 
is chosen on the basis of the signal-to-background ratio for 
the $K^0_S$ signal.  
A distance of about 10~cm yields a 
good background suppression.
Fig.~\ref{fig:armenteros} shows the Armenteros plot of the hadron 
pair, where the transverse momentum $p_T$ of one of the hadrons 
relative to the hadron momentum sum is plotted vs.\ the difference 
of the longitudinal momenta over their sum 
$(p_{L1}- p_{L2})/ (p_{L1}+ p_{L2} )$.  
In this plot the $K^0_S$ 
band is clearly seen, as well as the $\Lambda$ and $\bar{\Lambda}$ 
bands.  
To reduce the background due to $e^+ e^-$ pairs a lower cut 
on $p_{T}$ of 25~MeV is applied.
To finally identify the $K^0_S$ candidates a cut on
the invariant mass is applied. 
The reconstructed
invariant mass is required to be within $\pm 20$~MeV
of the known mass as shown in Fig.~\ref{fig:m_inv}. 
Since the width
of the fitted peak is $\sigma \approx 6$~MeV, the region
of $\pm 20$~MeV covers more than 99~\% of the signal.
Finally, the kaon transverse momentum 
with respect to the 
virtual photon direction is required
 to be larger than 0.1~GeV/$c$ to assure a good resolution in the measured 
azimuthal angle.
After all these cuts, the signal-to-background ratio is about 15,
constant over all the kinematic range.

Very much as for the charged hadron case, both ``leading $K^0_S$'' and ``all
K$^0_S$'' samples have been considered. 
In Table  \ref{tab:stat} the final statistics used for the asymmetry 
evaluation is given
for all the periods.

\section{Extraction of the asymmetries}

The Collins and Sivers asymmetries for ``all'' and ``leading'' pions
and kaons have been  evaluated
separately in kinematical bins of $x$, $p_T^h$ and $z$
while the other two variables were integrated over.
The number of events $N_{j,k}^{\pm}$ in the upstream and downstream target 
cell $(k=u,d)$ for the two polarisations ($+,-$) 
in a given 
$\Phi_j$ bin ($j=C,S$ refers to Collins and Sivers) can be written as 
\begin{equation}
\label{rawCounting}
N_{j,k}^{\pm}=F_k^{\pm} n_k \sigma a_{j,k}^{\pm}(\Phi_j)
\cdot(1 \pm \epsilon_{j,k}^{\pm} \sin \Phi_j).
\end{equation}
Here $F$ is the integrated incident muon flux, $n$ the number of target particles, 
$\sigma$ the spin averaged cross-section and $a_j$ the product of angular 
acceptance and efficiency of the spectrometer.
The quantities $\epsilon_{j,k}^\pm$ are given by
\begin{equation}
\epsilon_{C,k}^\pm =f\cdot P_{T,k}^\pm \cdot D_{NN} \cdot A_{Coll} \; , \; \; \;
\epsilon_{S,k}^\pm =f\cdot P_{T,k}^\pm \cdot A_{Siv}\; ,
\end{equation}
where $A_{Coll}$ and $A_{Siv}$ are the ``Collins'' and ``Sivers'' asymmetries,
related to the transversity and Sivers PDF respectively.  
The quantity
$D_{NN}=(1-y)/(1-y+y^2/2)$ is the transverse spin transfer coefficient 
from the target quark to the struck quark, 
$f$ 
is the dilution factor and $P_{T,k}^\pm$ the absolute value of the 
polarisation of the target cells.  
The dilution factor has been evaluated taking into 
account the radiative corrections for hadronic events, and 
is taken as constant, $f=0.38$, known to 5\%.  
The target polarisation $P_{T,k}^\pm$ has been measured for each 
cell and for each period \cite{Ageev:2006da} and averages to 48\% with a relative 
error of 5\%.

From the measured  number of events $N_{j,k}^{\pm}$, the ratio products
\begin{equation}
A_j(\Phi_j)=\frac{N_{j,u}^+(\Phi_j)}{N_{j,u}^-(\Phi_j)}
\cdot \frac{N_{j,d}^+(\Phi_j)}{N_{j,d}^-(\Phi_j)} \, , \; \; j=C,S 
\end{equation}
are computed and fitted with the functions 
$p_0\cdot(1+\hat A_j\cdot \sin \Phi_j)$ to extract 
the raw Collins and Sivers asymmetries.
Due to the smallness of the asymmetries involved, $\hat A_j$ is to a good approximation
$\hat A_j=\epsilon_{j,u}^+ + \epsilon_{j,d}^+ +\epsilon_{j,u}^-
 +\epsilon_{j,d}^-$.
The fit is performed in the interval $(0,2\pi)$ which is divided into 16 bins
for charged hadrons and 8 bins for $K^0_S$ due to the low statistics.  
From Monte Carlo simulations it is found that the angular resolution is much 
better than the bin size.

The Collins and Sivers asymmetries have been evaluated 
separately in each kinematic bin and for each data taking period; 
the results for the different periods have been combined 
using the weighted mean.
Finally using the purities a purity matrix $P$ can be written
\begin{eqnarray}
\label{prm}
P =
\left(
\begin{array}{cc}
P_{\pi, \pi} & P_{K, \pi}  \\
P_{\pi,  K}  & P_{K, K  }  \\
\end{array} \right),
\end{eqnarray}
here $P_{\pi, \pi}$ ($P_{K,K}$) is the fraction of real pions (kaons) inside the 
identified pion (kaon) sample and $P_{\pi,K}$
($P_{K,\pi}$)
is the fraction of misidentified kaons (pions) inside the pion (kaon)
sample. Due to the fact that the contribution of other particles (like protons)
was found to be negligible it is assumed that $P_{\pi,K}= 1 - P_{\pi,\pi}$ and  
$P_{K,\pi} = 1 - P_{K,K}$. 
The relation between measured $A\mathrm{^m}$ and corrected $A\mathrm{^c}$
asymmetries is
\begin{equation}
\label{truea}
\begin{array}{rcl}
\vec{A\mathrm{^{c}}} & = & P^{-1} \vec{A\mathrm{^{m}}} 
\end{array}
\hspace{5em}\mathrm{with}\hspace{1em}
\vec{A} = \left( \begin{array}{c} A_\pi \\[1ex] A_K \end{array} \right).
\end{equation}
This gives for the kaons
\begin{equation}
\label{cork}
A\mathrm{^c_K}  = \frac{1}{P_{\pi,\pi} +  P_{K,K} -1 } \Big[ P_{\pi,\pi} A\mathrm{^m_K} -  (1 - P_{K,K}) A\mathrm{^m_\pi}  \Big]. \nonumber 
\end{equation}
This equation contains two terms; the first accounts for the dilution of the
kaon asymmetry due to the purity of the sample, the second removes the
contribution of the pion asymmetry from the identified kaons. Since both pion
and kaon asymmetries on the deuteron target are compatible with zero over the
full range, this second term has been put to zero, both in Eq.~(\ref{cork}) and in the
equation giving $A\mathrm{^c_\pi}$, and the errors have been evaluated
correspondingly.

The same approach was followed for the neutral kaons, where it has been checked
that the asymmetry of the background under the $K^0_S$ peak was compatible with
zero.

\begin{figure}[t!] 
\begin{center}
\includegraphics[width=0.80\textwidth]{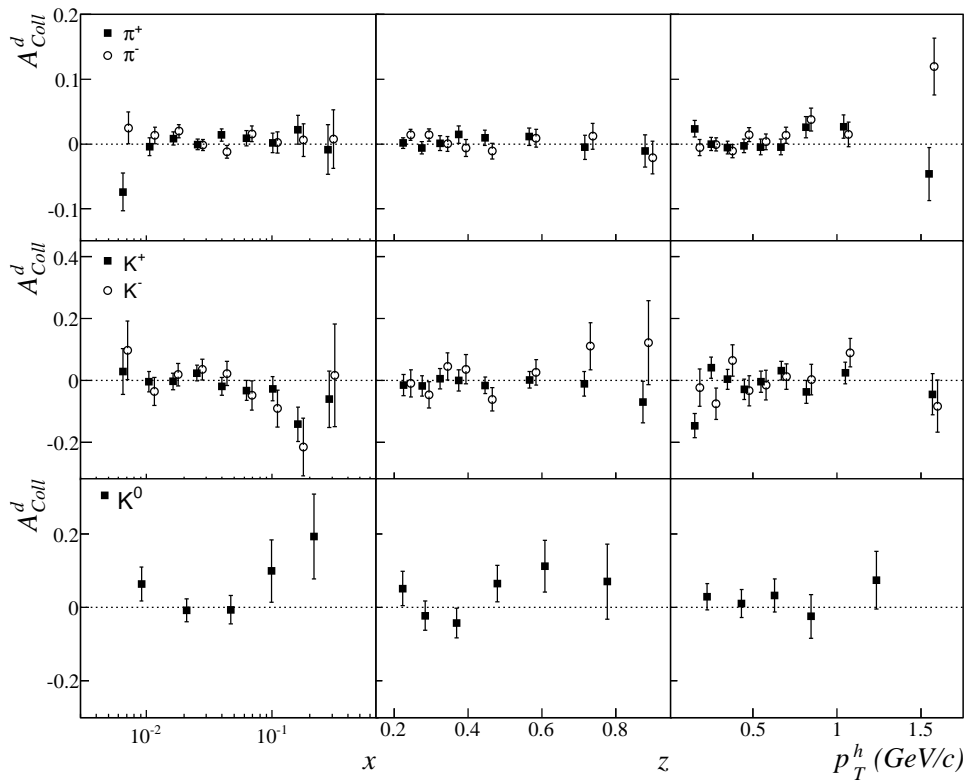}
\end{center}
\caption{Collins asymmetry against $x$, $z$ and $p_T^h$ for the ``all'' 
 charged pions and kaons samples from the 2003--2004 data, and the ``all'' $K^0_S$'s sample from the 2002--2004 data.}
\label{fig:r20034ac} 
\end{figure}
\begin{figure}[t!] 
\begin{center}
\includegraphics[width=0.80\textwidth]{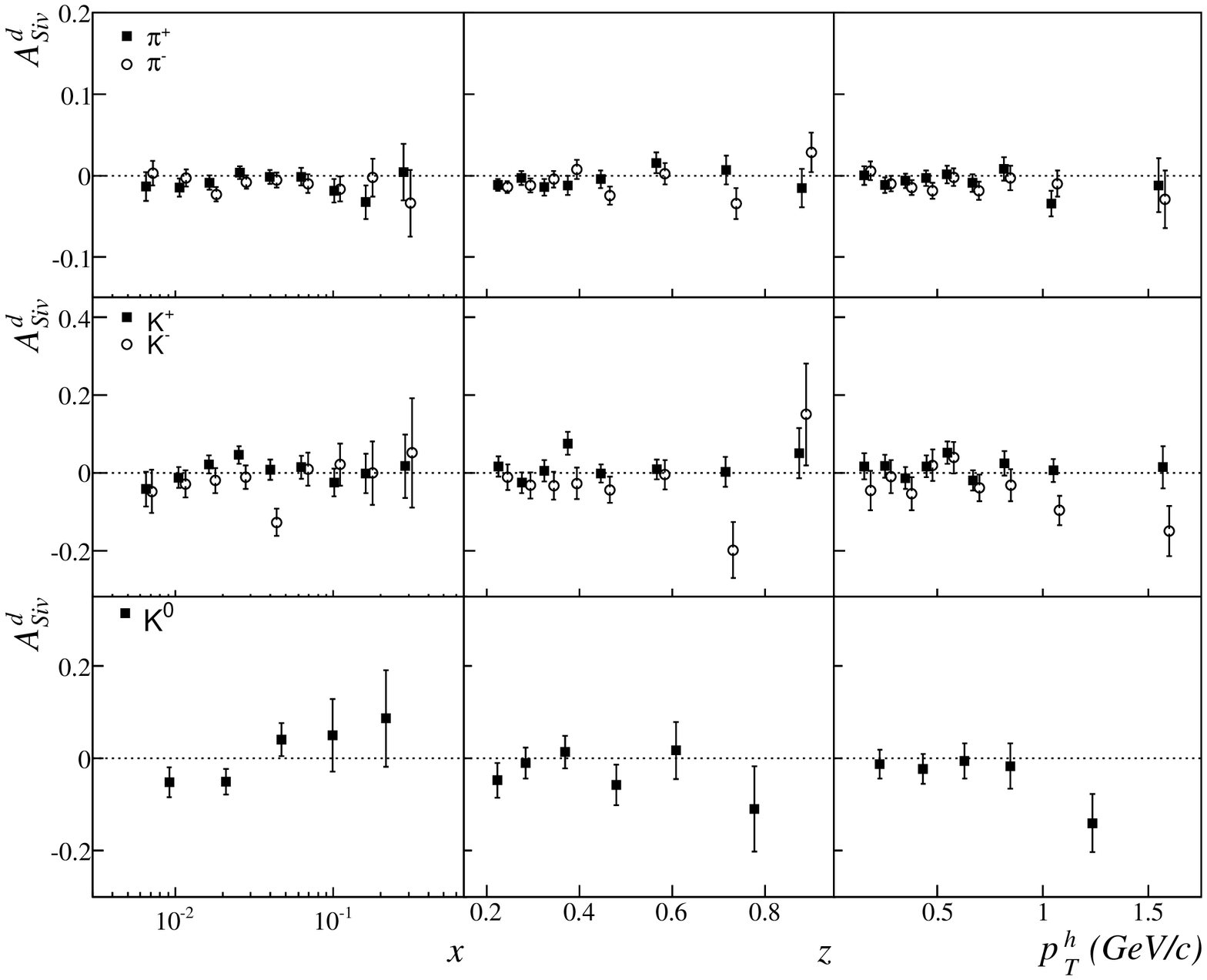}
\end{center}
\caption{Sivers asymmetry against $x$, $z$ and $p_T^h$ for the ``all'' 
 charged pions and kaons samples from the 2003--2004 data, and ``all'' $K^0_S$'s sample from the 2002--2004 data.}
\label{fig:r20034as} 
\end{figure}
\begin{figure}[t!] 
\begin{center}
\includegraphics[width=0.80\textwidth]{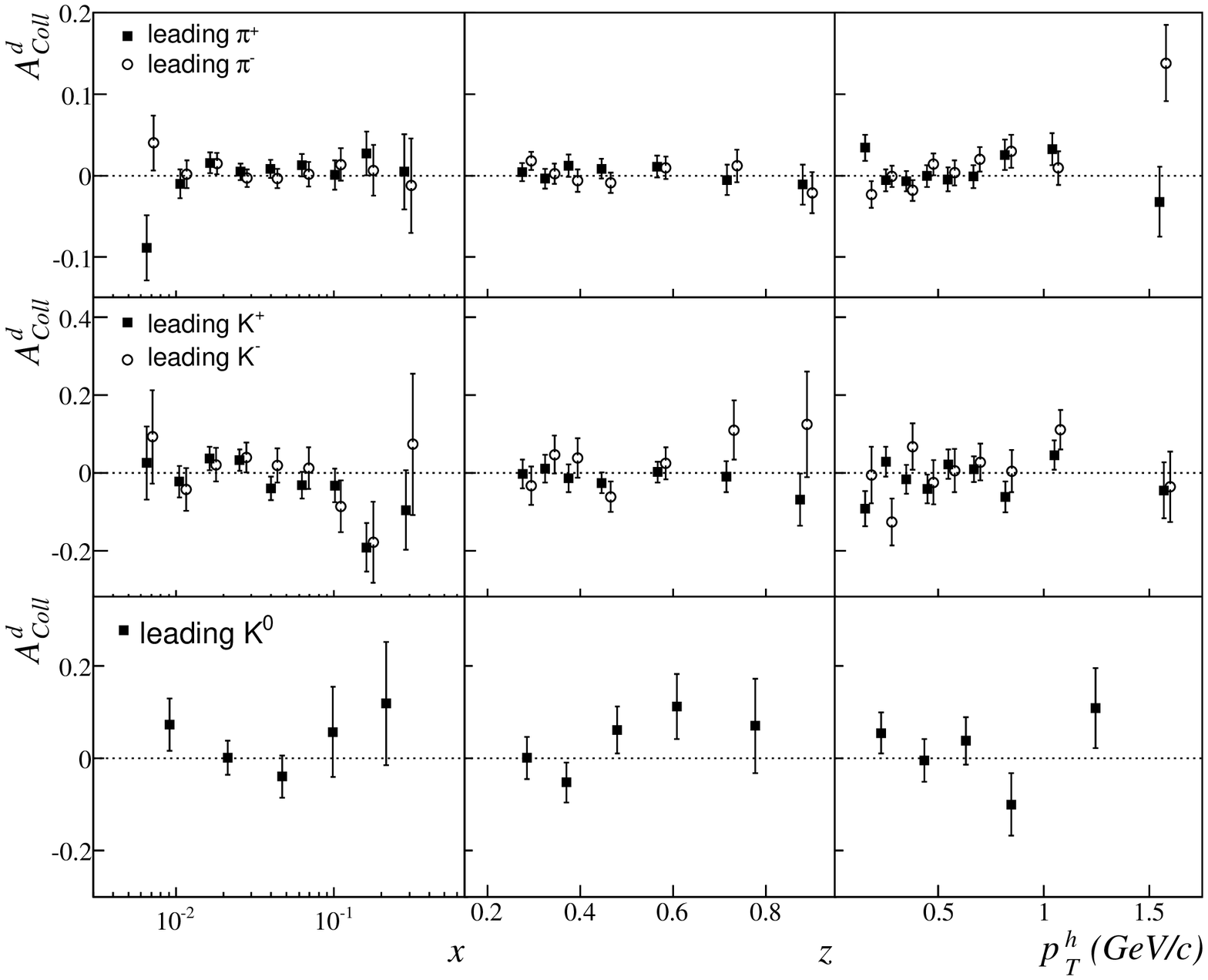}
\end{center}
\caption{Collins asymmetry against $x$, $z$ and $p_T^h$ for the ``leading'' 
 charged pions and kaons samples from the 2003--2004 data, and ``leading'' $K^0_S$'s sample from the 2002--2004 data.}
\label{fig:r20034lc} 
\end{figure}
\begin{figure}[t!] 
\begin{center}
\includegraphics[width=0.80\textwidth]{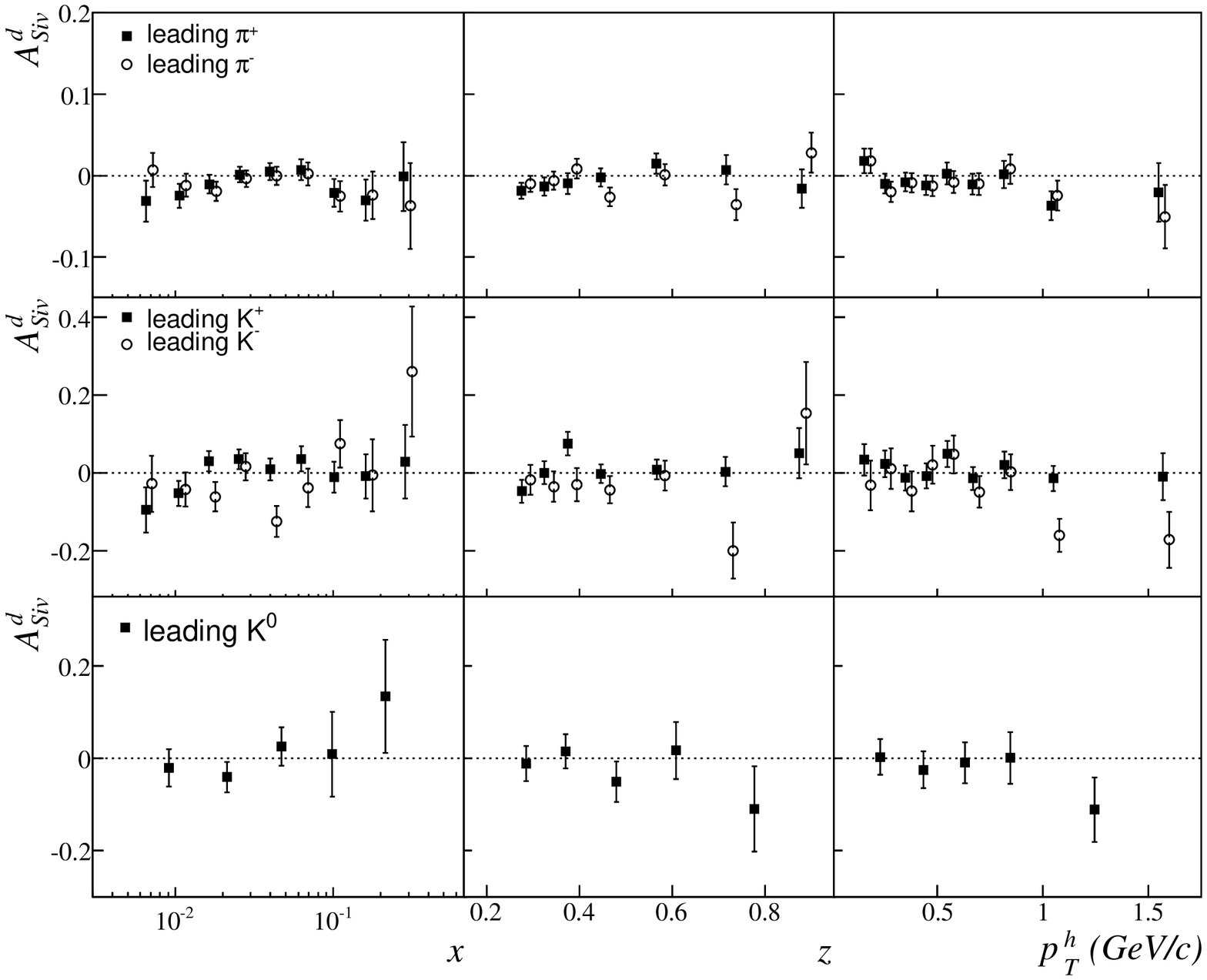}
\end{center}
\caption{Sivers asymmetry against $x$, $z$ and $p_T^h$ for the ``leading'' 
 charged pions and kaons samples from the 2003--2004 data, and ``leading'' $K^0_S$'s sample from the 2002--2004 data.}
\label{fig:r20034ls} 
\end{figure}

\section{Results and conclusion}

The final results for the Collins and Sivers asymmetries $A_{Coll}$ and
$A_{Siv}$ for charged pions and charged and neutral kaons on the deuteron target
vs.\ the three kinematic variables $x$, $z$ and $p_T^h$ are given in
Figs.~\ref{fig:r20034ac}--\ref{fig:r20034ls}.\footnote{ All the numerical 
values, including the purities, are available on HEPDATA.} In the figures,
the data points for negative hadrons, which are calculated in the 
same $x$-, $z$- and $p_T$-bin as for the positive hadrons, have been slightly 
shifted for graphical reasons. 



Extensive studies to evaluate the size of the systematic error have been 
performed. For some of these studies the $z$ cut has been opened and 
the data sample has been enlarged by a factor of three. 
The measured Collins and Sivers asymmetries were checked against 
stability among the five different periods of data taking, against
the use of different estimators to extract the asymmetries, 
against the reduction of the fiducial volume of our spectrometer
and against the influence of the trigger system of the experiment.
In these studies no deviations from the real asymmetries 
beyond the expected statistical fluctuations was observed. 
Furthermore experimental false asymmetries have been studied by combining 
the data set in such a way that the extracted asymmetries are expected to 
be zero. During all these tests no asymmetries
deviating from zero with statistical significance was observed.

%
Also, the correlation between the measured Collins and Sivers asymmetries which
originates from the non-uniform $\phi_h$/$\phi_S$ acceptance of the spectrometer
has been studied and the corresponding systematic error has been evaluated to be
negligible as compared with the statistical error.  The smallness of the
asymmetries makes the systematic error due to the uncertainties on $P_T$ and $f$
totally negligible.
These studies altogether lead to the final conclusion that 
the systematic errors are considerably smaller (well below 30\%) than the statistical errors. 
%

All the measured asymmetries are small, a trend which was already observed in
the published data of the non-identified hadrons. Small asymmetries are not a
surprise, it was expected that transverse spin effects be small in the deuteron
due to the opposite sign which was predicted for the u- and d-quark
distributions, very much like in the helicity case.

The interpretation of the results on the deuteron can be done only in
conjunction with corresponding proton data, measured by the HERMES Collaboration
albeit at lower energy.  Proton target data have been collected by COMPASS in
2007, but the results are not final at the time of writing.  As shown in
Refs.~\cite{Diefenthaler:2007rj,Ageev:2006da} a simple analysis of the HERMES
charged pion data and of the non-identified charged hadron data in COMPASS,
assuming that all the hadrons are pions, led to the following conclusions:
\begin{enumerate}
\item the favoured and unfavoured Collins functions have about the same size
    and the COMPASS deuteron data are needed for the
    extraction of the d-quark transversity;
\item  the null result for the Sivers asymmetry for the COMPASS data is a
clear indication that the u- and d-quark Sivers distribution
functions have about the same size and opposite sign.
\end{enumerate}
The same conclusions have been obtained in several analyses, 
using more sophisticated tools (see
e.g.~Refs.~\cite{Anselmino:2005ea,Vogelsang:2005cs,Efremov:2007kj}).  
A first global analysis 
which combined the 2002--2004 HERMES pion Collins asymmetries, the COMPASS 
results for non-identified hadrons, and the 
BELLE data has recently allowed 
to extract the Collins functions 
and, for the first time, the transversity 
distributions for the u- and d-quark~\cite{Anselmino:2007fs}.
Similar analyses can be now done including the present pion data which put more
stringent constraints.

The kaon data again show small asymmetries.  In the case of the charged kaons,
although the statistics is a factor of about 6 smaller 
than for the pions, the error bars are still rather small.  
The neutral kaon sample is smaller in size by a factor of about 3 
with respect to the charged kaons, and the error bars start being large.  
The COMPASS data do not exhibit the large difference between
$K^+$ and $\pi^+$ asymmetries seen by HERMES.  
Very much like for the $\pi^{\pm}$ case, cancellations are 
expected between u- and d-quarks when using the isoscalar deuteron target.  
Therefore the smallness of the COMPASS kaon asymmetries suggests that the 
sea quark contributions to the asymmetries are small.  
The kaon data provide a unique handle on the s-quark, but 
in this case the sea-quark contributions can not be 
neglected, and a full global analysis including pions and kaons is 
mandatory.

To summarise, COMPASS has made the first precise measurements of the Collins and
Sivers asymmetries for charged pions with a transversely polarised deuteron
target.  The same asymmetries have also been obtained for charged and neutral
kaons.  All the measured asymmetries are small, pointing at a cancellation
between the u- and d-quarks contributions.  More quantitative information, in
particular for the s-quark distributions, can be obtained with global analyses,
in which the COMPASS measurements with a transversely polarised proton target
undoubtedly will play an important role.

We acknowledge the support of the CERN management and staff, the special support
of CEA/Saclay in the target magnet project, as well as the skills and efforts of
the technicians of the collaborating institutes.

\end{document}